\documentclass[a4paper,12pt]{article}
\usepackage[utf8]{inputenc}
\usepackage[whole]{bxcjkjatype}

\catcode`\@=11
\@addtoreset{equation}{section}

\global\arraycolsep=2pt
\usepackage{amsmath}
\usepackage{amssymb}
\usepackage{mathtools}
\usepackage{graphicx}

\usepackage{comment}
\usepackage{cite}
\usepackage{physics}
\usepackage{subcaption}
\usepackage{bbm}
\begin{document}

\renewcommand{\thefootnote}{\arabic{footnote}}
\newcommand{\bhline}[1]{\noalign{\hrule height #1}}
\newcommand{\bvline}[1]{\vrule width #1}

\setcounter{footnote}{0}


\renewcommand{\thefootnote}{\fnsymbol{footnote}}

\begin{flushright}
KUNS-2942
\end{flushright}
\vspace*{1.5cm}

\begin{center}
{\Large \bf Chaotic string motion in a near pp-wave limit
}
\vspace*{2cm} \\
{\large Shodai Kushiro\footnote{E-mail:~kushiro$\_$at$\_$gauge.scphys.kyoto-u.ac.jp}
and Kentaroh Yoshida\footnote{E-mail:~kyoshida$\_$at$\_$gauge.scphys.kyoto-u.ac.jp}} 
\end{center}

\vspace*{0.4cm}

\begin{center}
{\it Department of Physics, Kyoto University, Kyoto 606-8502, Japan.}
\end{center}

\vspace{2cm}

\begin{abstract}
We revisit classical string motion in a near pp-wave limit of AdS$_5\times$S$^5$\,. It is known that the Toda lattice models are integrable.  
But if the exponential potential is truncated at finite order, then the system may become non-integrable. In particular, when the exponential potential in a three-particle periodic Toda chain is truncated at the third order of the dynamical variables, the resulting system becomes a well-known non-integrable system,  Henon-Heiles model. The same thing may happen in a near pp-wave limit of AdS$_5\times$S$^5$, on which the classical string motion becomes chaotic. 
\end{abstract}

\setcounter{footnote}{0}
\setcounter{page}{0}
\thispagestyle{empty}

\newpage

\tableofcontents
\renewcommand\thefootnote{\arabic{footnote}}

\section{Introduction}

The AdS/CFT correspondence\cite{Maldacena:1997re,Gubser:1998bc,Witten:1998qj} is one of the most significant subjects in String Theory. A typical example is the duality between type IIB superstring on AdS$_5\times$S$^5$ and the $\mathcal{N}=4$ SU($N$) super Yang-Mills (SYM) theory in the large $N$ limit. This duality has a nice property that both theories have an integrable structure \cite{Beisert:2010jr}. On the string-theory side, it is realized as classical integrability of the string sigma model whose target spacetime is given by AdS$_5\times$S$^5$ \cite{Mandal:2002fs,Bena:2003wd}. On the $\mathcal{N}=4$ SYM side, the integrable structure appears in the dilatation operator \cite{Minahan:2002ve}. For example, the one-loop planar dilatation operator in the SU(2) sector is identical to the Hamiltonian of the Heisenberg spin chain. 

\medskip 

This integrability may be broken for some cases. For the dilation operator on the gauge-theory side, for example, when non-planar corrections are included, the distribution of the energy-level spacings is close to the Wigner-Dyson distribution that indicates quantum chaos \cite{McLoughlin:2020siu,McLoughlin:2020zew}. More strikingly, the integrability may also be broken even in the large $N$ limit. When all-loop integrable spin chain \cite{Serban:2004jf,Beisert:2004hm,Beisert:2005fw} is truncated to finite order, the resulting system may exhibit quantum chaos \cite{McLoughlin:2022jyt}. That is, even if the full-order system is integrable, a finite-order truncation may lead to a non-integrable system. 


\medskip 

It is instructive to remember the result of \cite{McLoughlin:2022jyt}. When the full-order dilatation operator is truncated at two-loop level, the resulting system is described as a Heisenberg spin chain Hamiltonian with a next-nearest-neighbor (NNN) interaction: 
\begin{align}
H=\epsilon_0\sum^L_{i=1}\mathbbm{1}_{i,i+1}+\lambda_1\sum^L_{i=1}\mathbf{S}_{i}\cdot\mathbf{S}_{i+1}+\lambda_2\sum^L_{i=1}\mathbf{S}_{i}\cdot\mathbf{S}_{i+2}
\end{align}
with
\begin{align}
    \epsilon_0=1-\frac{3\lambda}{16\pi^2}\,,\qquad\lambda_1=-4+\frac{\lambda}{\pi^2}\,,\qquad\lambda_2=-\frac{\lambda}{4\pi^2}\,, 
\end{align}
where $\lambda$ is the 't Hooft coupling and the spin variable at the $i$-th site is represented by $\mathbf{S}_i=(\sigma^1_i, \sigma^2_i, \sigma^3_i)$\,, where $\sigma^a_i~(a=1,2,3)$ are the standard Pauli matrices. It is shown that the energy-level spacings of this system are well approximated by the Wigner-Dyson distribution and indeed this Hamiltonian indicates quantum chaos at finite $\lambda$\,. 

\medskip 

Motivated by the result of \cite{McLoughlin:2022jyt}, it is intriguing to study the string-theory side. One may consider a Penrose limit \cite{Penrose1,Penrose2} of the AdS$_5\times$S$^5$ background. When the light-cone coordinates are composed of the global AdS time and an angular variable in S$^5$\,, the resulting background becomes the maximally supersymmetric pp-wave background \cite{BFHP1,BFHP2}. The type IIB superstring theory on the pp-wave background is exactly solvable because the world-sheet theory is realized as a collection of massive free boson and fermion theories \cite{Metsaev}. The AdS/CFT dictionary at the pp-wave level was revealed by Berenstein, Maldacena and Nastase \cite{Berenstein:2002jq}. It is also possible to include sub-leading corrections in the Penrose limit. In this near pp-wave limit, the world-sheet theory has non-trivial interactions \cite{Callan:2003xr}. 

\medskip 

An intriguing question is whether the world-sheet theory in the near pp-wave limit is classically integrable or not. Indeed, this was studied in \cite{Asano:2015qwa}, where classical chaos has not been shown in this limit. However, the analysis there was based on a point particle ansatz, and the setup was really restricted. In this note, we will revisit this question with a stringy ansatz and show that the motion exhibits classical chaos by computing Poincar\'{e} sections and Lyapunov spectrum. As a result, the pp-wave string and the AdS$_5\times$S$^5$ string are integrable but the world-sheet theory in the middle of the Penrose limit is not integrable. 

\medskip

This result is quite analogous to the Henon-Heiles model, which is realized as a third-order truncation of a three-particle periodic Toda chain. It is well known that the Toda chain is integrbale but the Henon-Heiles model is not.

\medskip 

This note is organized as follows. Section 2 shows that the Henon-Heiles model is obtained by truncating a periodic Toda chain with three particles at third order of the dynamical variables. This is a typical example that a truncated system becomes chaotic. In section 3, we present the light-cone Hamiltonian in a near pp-wave limit of AdS$_5\times$S$^5$\,. In section 4, a reduced system is derived under a stringy ansatz. Then we compute Poincar\'{e} sections and Lyapunov spectrum. Section 5 is devoted to conclusion and discussion. Appendix A explains the derivation of the third conserved charge in the Toda chain. 


\section{Breaking classical integrability by truncation}

It is interesting to see a simple example that classical integrability is broken via a finite-order truncation of an integrable system. Here, let us consider a three-particle periodic Toda chain and derive a non-integrable Henon-Heiles model as a finite-order truncation.

\medskip

\subsection{Three-particle periodic Toda chain}

The Hamiltonian of a three-particle periodic Toda chain is given by 
\begin{align}
  H=\frac{1}{2}\qty(P_1^2+P_2^2+P_3^2)+e^{-\qty(Q_2-Q_1)}+e^{-\qty(Q_3-Q_2)}+e^{-\qty(Q_1-Q_3)}-3\,. 
  \label{expHam}
\end{align}
This system is classically integrable in the sense of Liouville. 
Hamilton's equations are
\begin{align}
  \dv{Q_1}{t}&=P_1\,,\qquad\dv{Q_2}{t}=P_2\,,\qquad\dv{Q_3}{t}=P_3\,,\\
  \dv{P_1}{t}&=e^{Q_3-Q_1}-e^{Q_1-Q_2}\,,\quad\dv{P_2}{t}=e^{Q_1-Q_2}-e^{Q_2-Q_3}\,,\quad\dv{P_3}{t}=e^{Q_2-Q_3}-e^{Q_3-Q_1}\,. 
  \label{eom-P}
\end{align}
The number of degrees of freedom is three and hence three conserved charges (in involution) ensure the classical integrability. 

\medskip

Let us see that the system (\ref{expHam}) has three conserved charges below. The first one is the total momentum $I_1 \equiv P_1+P_2+P_3$\,. By using (\ref{eom-P})\,, it is conserved as 
\begin{align}
  \dv{I_1}{t}=\dv{t}\qty(P_1+P_2+P_3)=0\,.
\end{align}
Next, in order to derive the second one, let us evaluate the time derivative of $P_1 P_2$ as follows: 
\begin{align}
  \dv{\qty(P_1P_2)}{t}&=P_1\dv{P_2}{t}+P_2\dv{P_1}{t}\nonumber\\
  &=\dv{Q_1}{t}\qty(e^{Q_1-Q_2}-e^{Q_2-Q_3})+\dv{Q_2}{t}\qty(e^{Q_3-Q_1}-e^{Q_1-Q_2})\nonumber\\
  &=\dv{t}e^{Q_1-Q_2}-\dv{Q_1}{t}e^{Q_2-Q_3}+\dv{Q_2}{t}e^{Q_3-Q_1}\,. \label{eq1}
\end{align}
Similar results can also be derived as 
\begin{align}
  \dv{\qty(P_2P_3)}{t}&=\dv{t}e^{Q_2-Q_3}-\dv{Q_2}{t}e^{Q_3-Q_1}+\dv{Q_3}{t}e^{Q_1-Q_2}\,, 
  \label{eq2}\\
  \dv{\qty(P_3P_1)}{t}&=\dv{t}e^{Q_3-Q_1}-\dv{Q_3}{t}e^{Q_1-Q_2}+\dv{Q_1}{t}e^{Q_2-Q_3}\,. 
  \label{eq3}
\end{align}
Then, by summing up the above three expressions (\ref{eq1})-(\ref{eq3}), one can see that 
\begin{align}
  \dv{t}\qty(P_1P_2+P_2P_3+P_3P_1-e^{Q_1-Q_2}-e^{Q_2-Q_3}-e^{Q_3-Q_1})=0 
\end{align}
and find the second conserved charge 
\begin{eqnarray}
I_2 \equiv P_1P_2+P_2P_3+P_3P_1
-e^{Q_1-Q_2}-e^{Q_2-Q_3}-e^{Q_3-Q_1}\,.  
\end{eqnarray}
The Hamiltonian \eqref{expHam} can be expressed as $H=I_1^2/2 - I_2 +3$ and thus $I_2$ is equivalent to the total energy.

\medskip

The last is to find out the third one. As explained in Appendix A in detail, after some algebra, the time derivative of $P_1P_2P_3$ can be rewritten as follows: 
\begin{align}
  \dv{t}\qty(P_1P_2P_3)=\dv{t}\qty(P_1e^{Q_2-Q_3}+P_2e^{Q_3-Q_1}+P_3e^{Q_1-Q_2})\,.
\end{align}
Thus $I_3=P_1P_2P_3-P_1e^{Q_2-Q_3}-P_2e^{Q_3-Q_1}-P_3e^{Q_1-Q_2}$ is the third conserved charge. As one can easily check, $I_1$, $I_2$ and $I_3$ are in involution. Hence this system is classically integrable in the sense of Liouville.

\medskip 

\subsection{Derivation of the Henon-Heiles model}

Here, let us derive the Henon-Heiles model as a third-order truncation of the Toda chain introduced in the previous subsection. 

\medskip 

Assume that $Q_i$'s are small and expand the exponential potentials in the Hamiltonian \eqref{expHam} in third order of $Q_i$\,. Then the resulting Hamiltonian is given by 
\begin{align}
  H&=H_0+H_1\,,\label{nlHam}\\
  H_0&\equiv \frac{1}{2}\qty(P_1^2+P_2^2+P_3^2)+\frac{1}{2}\qty{\qty(Q_1-Q_2)^2+\qty(Q_2-Q_3)^2+\qty(Q_3-Q_1)^2}\,,\\
  H_1&\equiv\frac{1}{6}\qty{\qty(Q_1-Q_2)^3+\qty(Q_2-Q_3)^3+\qty(Q_3-Q_1)^3}\,, 
\end{align}
where $H_0$ is the part of coupled harmonic oscillators and $H_1$ is the interaction term.

\medskip

It it convenient to perform a rotation represented by
\begin{align}
  Q_i=\sum^3_{j=1}A_{ij}\,\zeta_j\,,\qquad P_i=\sum^3_{j=1}A_{ij}\,\eta_j\,,
\end{align}
where the constant orthogonal matrix $A_{ij}$ ($A^{T}A=I$) is defined as 
\begin{align}
  A \equiv \mqty(\frac{1}{\sqrt6}&\frac{1}{\sqrt2}&\frac{1}{\sqrt3}\\
  -\frac{\sqrt{2}}{\sqrt{3}}&0&\frac{1}{\sqrt3}\\
  \frac{1}{\sqrt6}&-\frac{1}{\sqrt2}&\frac{1}{\sqrt3})\,.
\end{align}
Then the resulting Hamiltonian is given by 
\begin{align}
  H=\frac{1}{2}\qty(\eta_1^2+\eta_2^2+\eta_3^2+3\zeta_1^2+3\zeta_2^2)+\frac{3}{2\sqrt{2}}\qty(\zeta_2\zeta_1^2-\frac{1}{3}\zeta_2^3)\,.
\end{align}
Since the total momentum $P_1+P_2+P_3=\sqrt{3}\,\eta_3$ is conserved, we can drop off $\eta_3$ by setting that $\eta_3=0$\,.
Then Hamilton's equations are rewritten in terms of $\zeta_1$\,, $\zeta_2$\,, $\eta_1$ and $\eta_2$ as follows: 
\begin{align}
  \dv{\zeta_1}{t}&=\eta_1\,,\quad\dv{\zeta_2}{t}=\eta_2\,,\quad\dv{\eta_1}{t}=-3\zeta_1-\frac{3}{\sqrt{2}}\zeta_1\zeta_2\,,\quad\dv{\eta_2}{t}=-3\zeta_2-\frac{3}{2\sqrt{2}}\qty(\zeta_1^2-\zeta_2^2)\,.\label{eomlattice}
\end{align}
After rescaling the variables like 
\begin{align}
  q_1=\frac{1}{2\sqrt{2}}\,\zeta_1\,,\quad q_2=\frac{1}{2\sqrt{2}}\,\zeta_2\,,\quad p_1=\frac{1}{2\sqrt{6}}\,\eta_1\,,\quad p_2=\frac{1}{2\sqrt{6}}\,\eta_2\,,\quad \tau=\sqrt{3}\,t\,,
\end{align}
Hamilton's equations in \eqref{eomlattice} and the resulting Hamiltonian are given by, respectively, 
\begin{align}
  \dv{q_1}{\tau}=p_1\,,\quad &\dv{q_2}{\tau}=p_2\,,\quad \dv{p_1}{\tau}=-q_1-2q_1q_2\,,\quad \dv{p_2}{\tau}=-q_2-q_1^2+q_2^2\,,\\ 
  &H =\frac{1}{2}\qty(p_1^2+p_2^2)+\frac{1}{2}\qty(q_1^2+q_2^2)+q_2q_1^2-\frac{1}{3}q_2^3\,.
\end{align}
The system described by the above $H$ is known as the Henon-Heiles model.

\medskip

\begin{figure}[htbp]
\vspace*{1cm}
  \centering
  \includegraphics[width=7cm]{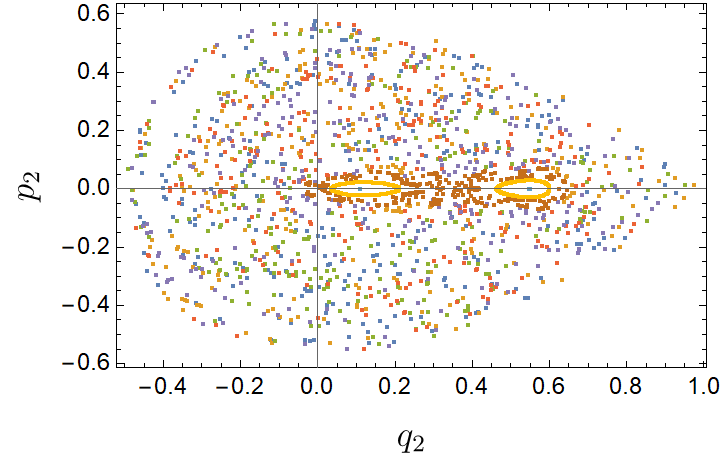}
  \caption{\footnotesize Poincar\'{e} section of the Henon-Heiles model with $E=1/6$}
  \label{fig:lattice}
\end{figure}

This model is well-known as a non-integrable model. A typical Poincar\'{e} section is depicted in Fig.\ \ref{fig:lattice}, where $E=1/6$\,, $q_1=0$ and $p_1>0$\,. Different colors mean different trajectories corresponding to different initial values. This result indicates that the motion becomes chaotic for almost all of the initial values. 

\medskip

Note here that the total momentum $I_1$ and the total energy (which is equivalent to $I_2$) remain to be conserved charges by the invariance of the system under spatial and time translations, while the third conserved charge does not exist. Actually, the expanded form of $I_3$ is not conserved. For the detail of this point, see Appendix \ref{appendix}.

\section{The Hamiltonian of a near pp-wave string}

In this section, we present the light-cone Hamiltonian of a bosonic string theory propagating on a near pp-wave limit of the AdS$_5\times$S$^5$ background by following the pioneering work \cite{Callan:2003xr}.

\subsection{A near pp-wave limit of AdS$_5\times$S$^5$}

Let us start from the metric of AdS$_5\times$S$^5$ with the global coordinates:
\begin{align}
  ds^2=R^2(-\cosh^2\rho\, dt^2+d\rho^2+\sinh^2\rho\,d\Omega^2_3+\cos^2\theta\, d\phi^2+d\theta^2+\sin^2\theta\, d\Omega^{\prime2}_3)\,.
  \label{adsmetric}
\end{align}
Here $d\Omega^2_3$ and $d\Omega^{\prime2}_3$ are the metrics of S$^3$'s, 
and $R$ is the common radius of $\mathrm{AdS}_5$ and S$^5$\,. Then the metric \eqref{adsmetric} can be rewritten as
\begin{align}
  ds^2=R^2\qty(-\qty(\frac{1+\tilde{z}^2/4}{1-\tilde{z}^2/4})dt^2+\qty(\frac{1-\tilde{y}^2/4}{1+\tilde{y}^2/4})d\phi^2
  +\frac{d\tilde{z}^2+\tilde{z}^2d\Omega^2_3}{\qty(1-\tilde{z}^2/4)^2}
  +\frac{d\tilde{y}^2+\tilde{y}^2d\Omega^{\prime2}_3}{\qty(1+\tilde{y}^2/4)^2})
\end{align}
by using the new coordinates $\tilde{z}$ and $\tilde{y}$ defined as  
\begin{align}
  \cosh\rho \equiv \frac{1+\tilde{z}^2/4}{1-\tilde{z}^2/4}\,,\qquad \cos\theta \equiv \frac{1-\tilde{y}^2/4}{1+\tilde{y}^2/4}\,.
\end{align}
By taking the light-cone coordinates as 
\begin{align}
  \tilde{x}^+=t\,,\qquad \tilde{x}^-=-t+\phi\,, 
\end{align}
and rescaling the coordinates as 
\begin{align}
  \tilde{x}^+ \longrightarrow x^+\,,\qquad \tilde{x}^-\longrightarrow \frac{x^-}{R^2}\,,\qquad \tilde{z}\longrightarrow \frac{z}{R}\,,\qquad \tilde{y}\longrightarrow \frac{y}{R}\,,
\end{align}
one may take the $R\to \infty$ limit. This is nothing but the Penrose limit \cite{Penrose1,Penrose2}. 
The resulting metric with the subleading correction \cite{Callan:2003xr} is given by
\begin{align}
  ds^2&=ds_0^2+\frac{1}{R^2}ds_2^2+\mathcal{O}\left(\frac{1}{R^4}\right)\,,\label{ppwave}\\
  ds_0^2& \equiv2dx^+dx^--(z^2+y^2)(dx^+)^2+dz^2+z^2d\Omega_{3}^{2}+dy^2+y^2d\Omega_{3}^{\prime2}\,,\label{leading}\\
  ds_2^2& \equiv -2y^2dx^+dx^-+\frac{1}{2}(y^4-z^4)(dx^+)^2+(dx^-)^2\nonumber\\
  &\qquad\qquad+\frac{1}{2}z^2(dz^2+z^2d\Omega_{3}^{2})-\frac{1}{2}y^2(dy^2+y^2d\Omega_{3}^{\prime2})\label{subleading}\,.
\end{align}
The leading term \eqref{leading} is the metric part of the maximally supersymmetric pp-wave background \cite{BFHP1,BFHP2}.

\subsection{Light-cone Hamiltonian} 

Next, we will derive the light-cone Hamiltonian of a string theory on the near pp-wave background \eqref{ppwave}. For simplicity, we focus on the bosonic sector. The subscripts $\mu,\nu$ denote the ten-dimensional indices $+,-,1,\dots, 8$ and $I,J$ denote only the transverse directions $1,\dots,8$\,.

\medskip

The classical string action is given by
\begin{align}
  S=\int d\tau d\sigma \mathcal{L}=\frac{1}{2}\int d\tau d\sigma \sqrt{-\det (h_{ab})} h^{ab}\partial_a x^{\mu}\partial_b x^{\nu} g_{\mu\nu}\,,
\end{align}
where $g_{\mu\nu}$ is the background metric \eqref{ppwave} and $x^{\mu}$ are the target-space coordinates. The world-sheet metric is given by $h_{ab}~(a,b=\tau,\sigma)$\,. The equation of motion for $h_{ab}$ imposes that the stress-energy tensor $T_{ab}$ on the world-sheet should vanish and lead to 
the Virasoro constraints: 
\begin{align}
  T_{ab}=\partial_a x^{\mu}\partial_b x^{\nu}g_{\mu\nu}-\frac{1}{2}h_{ab}h^{cd}\partial_c x^{\mu} \partial_d x^{\nu} g_{\mu\nu}=0\,. 
  \label{virasoro} 
\end{align}
The canonical momenta and their inversion in terms of $\dot{x}^{\mu}$ are given by 
\begin{align}
  p_{\mu}=\pdv{\mathcal{L}}{(\partial_{\tau}x^{\mu})}=h^{\tau a}\partial_ax^{\nu}g_{\mu\nu}\,,\qquad \dot{x}^{\mu}=\frac{1}{h^{\tau\tau}}g^{\mu\nu}p_{\nu}-\frac{h^{\tau\sigma}}{h^{\tau\tau}}x^{\prime\mu}\,,
\end{align}
where $\dot{x}^{\mu}\equiv\partial x^{\mu}/\partial\tau$ and $x^{\prime\mu}\equiv\partial x^{\mu}/\partial\sigma$.
Then the Virasoro constraints \eqref{virasoro} can be rewritten as
\begin{align}
  p_{\mu}p_{\nu}g^{\mu\nu}+x^{\prime\mu}x^{\prime\nu}g_{\mu\nu}&=0\,,\label{virasoro1}\\
  p_{\mu}x^{\prime\mu}&=0\,.
\end{align}

\medskip

Here, let us employ the light-cone gauge:
\begin{align}
  x^{+}=\tau\,,\qquad p_{-}=\mathrm{const}\,.
\end{align}
Then the light-cone Hamiltonian is defined as
\begin{align}
  \mathcal{H}_{\mathrm{lc}}\equiv -p_{+}\,.
\end{align}
By solving the equation \eqref{virasoro1} in terms of $\mathcal{H}_{\mathrm{lc}}$\,, the light-cone Hamiltonian is expressed as 
\begin{align}
  \mathcal{H}_{\mathrm{lc}}=\frac{p^{-}g^{+-}}{g^{++}}-\frac{1}{g^{++}}\sqrt{p_{-}^2g-g^{++}\qty(g_{--}\qty(\frac{p_{I}x^{\prime I}}{p_{-}^2})^2+p_I p_J g^{IJ}+x^{\prime I}x^{\prime J}g_{IJ})}\,,
  \label{lightcone}
\end{align}
where 
\begin{align} 
  g\equiv (g^{+-})^2-g^{++}g^{--}\,.
\end{align}

\medskip

Finally, we write down the light-cone Hamiltonian on the near pp-wave background (\ref{ppwave})\,. Let  $(g_{0})_{\mu\nu}$ and $(g_2)_{\mu\nu}$ denote the leading term \eqref{leading} and the sub-leading term \eqref{subleading}, respectively. Then the background metric is expressed as 
\begin{align}
  g_{\mu\nu}=(g_{0})_{\mu\nu}+\frac{1}{R^2}(g_2)_{\mu\nu}+\mathcal{O}\qty(\frac{1}{R^4})\,,\qquad g^{\mu\nu}=(g^{\prime}_0)^{\mu\nu}+\frac{1}{R^2}(g_2^{\prime})^{\mu\nu}+\mathcal{O}\qty(\frac{1}{R^4})\,,
  \label{metric}
\end{align}
where $g_0^{\prime}=g_0^{-1}$ and $g_2^{\prime}=-g^{-1}_0g_2g^{-1}_0$.
By substituting the expanded metric \eqref{metric} for \eqref{lightcone}, the light-cone Hamiltonian \cite{Callan:2003xr} is evaluated as 
\begin{align}
  \mathcal{H}_{\mathrm{lc}}&=\mathcal{H}_0+\frac{1}{R^2}\mathcal{H}_{\mathrm{int}}+\mathcal{O}\qty(\frac{1}{R^4})\,,\label{abababc}\\
  \mathcal{H}_0&=\frac{1}{2}\qty(p_I p_J(g^{\prime}_0)^{IJ}+x^{\prime I}x^{\prime J}(g_{0})_{IJ}+y^2+z^2)\,,\\
  \mathcal{H}_{\mathrm{int}}&=\frac{1}{8}\qty(\qty(y^2+z^2)^2-\qty(p_I p_J(g^{\prime}_0)^{IJ}+x^{\prime I}x^{\prime J}(g_{0})_{IJ})^2)+\frac{1}{2}\qty(p_Ix^{\prime I})\nonumber\\
  &\qquad+\frac{1}{4}\qty(z^2-y^2)\qty(p_I p_J(g^{\prime}_0)^{IJ}+x^{\prime I}x^{\prime J}(g_{0})_{IJ})
  +\frac{1}{2}\qty(p_I p_J(g^{\prime}_2)^{IJ}+x^{\prime I}x^{\prime J}(g_{2})_{IJ})\,.
\end{align}
Here we have set $p_-=1$. 
In the following, we will not consider the higher-order terms with $\mathcal{O}(1/R^4)$\,.
The leading term $\mathcal{H}_0$ is a sum of harmonic oscillators and hence this part is obviously integrable. The subleading correction in the Penrose limit gives rise to the interaction Hamiltonian $\mathcal{H}_{\mathrm{int}}$ which contains the fourth-order terms of the canonical variables $(x^{I},p_{I})$\,. Thus the question we want to study below is whether this interacting dynamics is integrable or not.

\section{Chaotic dynamics} 

In this section, let us answer the question, whether this interacting dynamics described by the Hamiltonian (\ref{abababc}) is integrable or not. Actually, it was considered in the preceding work \cite{Asano:2015qwa}, where a point-particle ansatz was considered and chaotic dynamics was not found. In the following, we will take another stringy ansatz and show classical chaos directly by computing Poincar\'{e} sections and Lyapunov spectrum.

\subsection{A stringy ansatz}
\label{sec:ansatz}

There are two S$^3$ parts in the metric (\ref{ppwave})\,. We suppose that the string is sitting at a point in S$^3$ from AdS$_5$ and hence the terms with $d\Omega^2_3$ are ignored below. 

\medskip

It is convenient to parametrize S$^3$ from S$^5$ by using the Hopf coordinates $(\eta,\xi_1,\xi_2)$~: 
\begin{align}
  d\Omega_3^{\prime2}=d\eta^2+\sin^2\eta\, d\xi_1^2+\cos^2\eta\, d\xi_2^2\,.
\end{align}
Note that $p_{\xi_1}$ and $p_{\xi_2}$ are conserved and hence we will set $p_{\xi_1}=p_{\xi_2}=0$ below. 

\medskip 

In order to study the chaotic dynamics, it is necessary to impose an ansatz so as to reduce the system from a two-dimensional field theory to a mechanical system. In the following, we employ a winding string ansatz~:
\begin{align}
  z=0\,,\quad p_z=0\,,\quad y=y(\tau)\,,\quad p_y=p_y(\tau)\,,\nonumber\\
  \quad \xi_1=a_1\sigma\,,\quad\xi_2=a_2\sigma\,\quad\eta=\eta(\tau)\,,\quad p_{\eta}=p_{\eta}(\tau)\,. \label{stringy}
\end{align}
Here $a_i~(i=1,2)$ are integers due to the periodicity of $\xi_i$\,. Note that this ansatz is consistent with the full equation of motion derived from $\mathcal{H}_{\mathrm{lc}}$ in (\ref{abababc})\,. 
This ansatz leads to the reduced equations of motion 
\begin{align}
\dot{y}&=p_y-\frac{1}{2R^2}\qty[P_y^3+\frac{P_yP_{\eta}^2}{y^2}+y^2\qty(a_1^2\sin^2\eta+a_2^2\cos^2\eta)]\,,\label{reduceom1}\\
\dot{\eta}&=\frac{p_{\eta}}{y^2}-\frac{1}{2R^2}\qty[\frac{p_{\eta}^3}{y^4}+\frac{p_y^2p_{\eta}}{y^2}+\qty(a_1^2\sin^2\eta+a_2^2\cos^2\eta)]\,,\label{reduceom2}\\
\dot{p_y}&=\frac{p_{\eta}^2}{y^3}-y\qty(a_1^2\sin^2\eta+a_2^2\cos^2\eta+1)\nonumber\\
& +\frac{1}{2R^2}\left[-5y^3-2yp_y^2-\frac{p_{\eta}^4}{y^5}-\frac{p_y^2p_{\eta}^2}{y^3}-\frac{p_y^4}{4y}+y^3\qty(a_1^2\sin^2\eta+a_2^2\cos^2\eta+\frac{p_y^2+4y^2}{2y^2})^2\right]\,,\label{reduceom3}\\
\dot{p_{\eta }}&=\frac{1}{2}\qty(a_2^2-a_1^2)y^2\sin 2\eta\nonumber \\ 
& -\frac{1}{8R^2}\qty(a_2^2-a_1^2)\sin 2\eta\qty[\qty(a_2^2-a_1^2)y^4\cos 2\eta+\qty(a_1^2+a_2^2+4)y^4+2p_{\eta}^2+2y^2p_y^2] \,. 
\label{reduceom4}
\end{align}
These can also be reproduced from the reduced light-cone Hamiltonian~:
\begin{align}
  \mathcal{H}_{\mathrm{lc}}&=\mathcal{H}_0+\frac{1}{R^2}\mathcal{H}_{\mathrm{int}}\,,\label{trunHam}\\
  \mathcal{H}_0&=\frac{1}{2}\qty(p_y^2+\frac{p_\eta^2}{y^2}+y^2+y^2(a_1^2\sin^2\eta+a^2_2\cos^2\eta))\,,\\
  \mathcal{H}_{\mathrm{int}}&=y^2p_y^2+p_{\eta}^2+y^4\qty(1-a_1^2\sin^2\eta-a_2^2\cos^2\eta) \notag \\ 
& \quad  -\frac{\qty(y^2p_y^2+p_\eta^2+y^4\qty(1+a_1^2\sin^2\eta+a_2^2\cos^2\eta))^2}{2y^4}\,.
\end{align} 
When $a_1^2=a_2^2$\,, the reduced system is independent of $\eta$\,. In the following analysis, we set $a_1=1$ and $a_2=2$ so as to keep the contribution of $\eta$\,. Moreover, the AdS radius $R$ is set to $R=5.0$ for simplicity.

\subsection{Poincar\'{e} sections and Lyapunov spectrum}

\begin{figure}[htbp]
  \vspace*{0.7cm}
  \centering
  \begin{tabular}{c}
   \begin{minipage}{0.5\hsize}
   \centering
   \includegraphics[width=7cm]{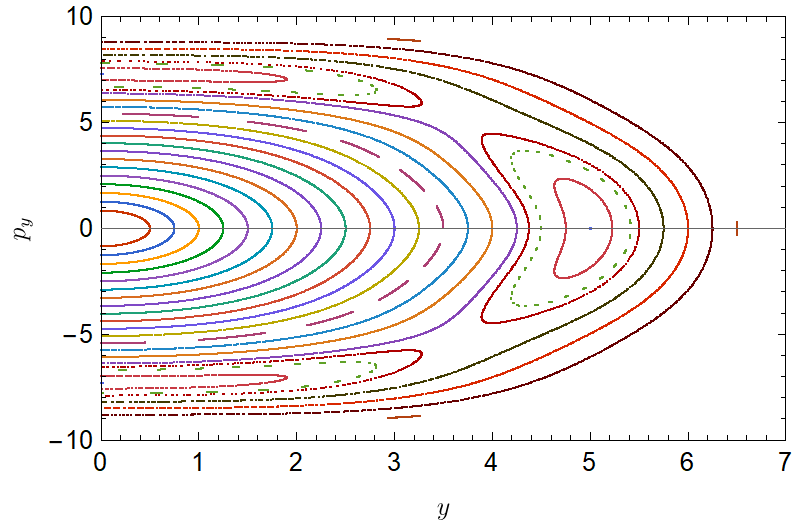}
   \subcaption{\footnotesize Poincar\'{e} section for $E=5.0$}
   \label{fig:t11a}
   \end{minipage}
   \begin{minipage}{0.5\hsize}
   \centering
   \includegraphics[width=7cm]{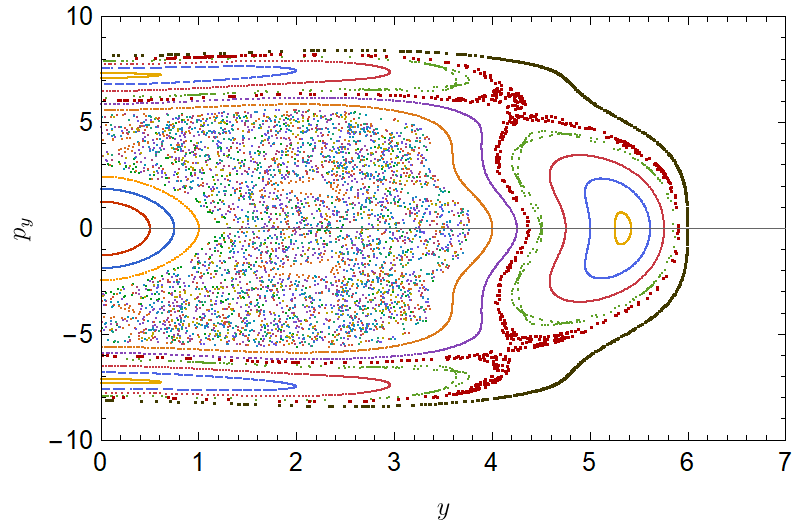}
   \subcaption{\footnotesize Poincar\'{e} section for $E=11.0$}
   \label{fig:t11b}
   \end{minipage}
   \vspace*{0.7cm}\\
   \begin{minipage}[b]{0.5\hsize}
   \centering
   \includegraphics[width=7cm]{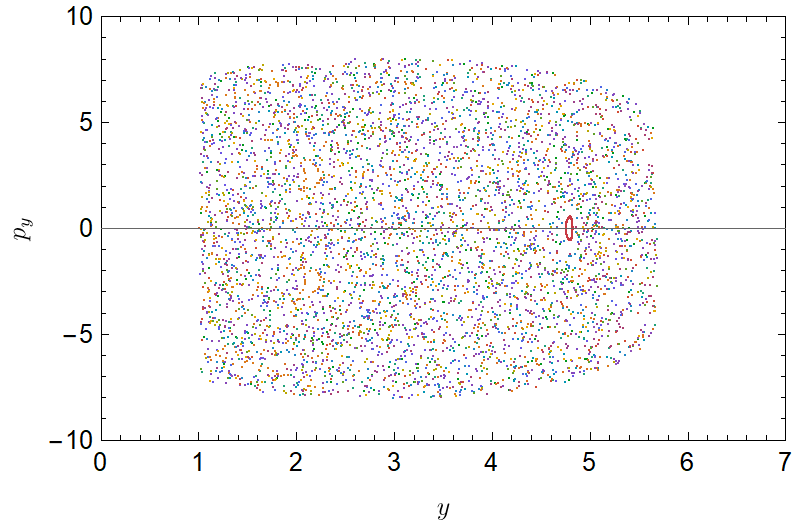}
   \subcaption{\footnotesize Poincar\'{e} section for $E=13.0$}
   \label{fig:t11c}
   \end{minipage}
   \begin{minipage}[b]{0.5\hsize}
   \centering
   \includegraphics[width=7cm]{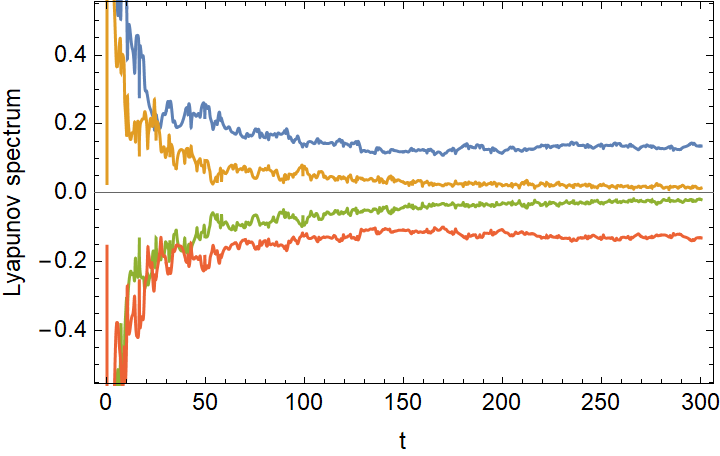}
   \subcaption{\footnotesize Lyapunov spectrum for  $E=13.0$}
   \label{fig:t11d}
   \end{minipage}
  \end{tabular}
  \caption{\footnotesize Poincar\'{e} sections and Lyapunov spectrum}
  \label{fig:2}
\end{figure}

We compute Poincar\'{e} sections and Lyapunov spectrum for the set of equations of motion (\ref{reduceom1})--(\ref{reduceom4})\,.
Poincar\'{e} sections are taken at $\eta=\pi/2$ with $p_{\eta}>0$ and plotted for $E=5.0,\,11.0,\,13.0$ as depicted in Figs.\,\ref{fig:2}\,(a)-(c), respectively. Different colors indicate different initial values satisfying the energy constraint. When $E=5.0$, there are no chaotic motions and only quasi-periodic orbits exist [Fig.\,\ref{fig:t11a}]. These periodic orbits are called the Kolmogorov-Arnold-Moser (KAM) tori \cite{Kolmogorov:1954,Arnold:1963,Moser:1962}. When $E=11.0$\,, some KAM tori have collapsed and classical chaos appears [Fig.\,\ref{fig:t11b}]. When $E=13.0$\,, almost all orbits become chaotic [Fig.\,\ref{fig:t11c}].

\medskip

Figure\,\ref{fig:t11d} is a Lyapunov spectrum for $E=13.0$\,. The initial values are taken as $y=2.0$\,, $\eta=\pi/2$ and $p_y=1.0$. Then the maximum Lyapunov exponent is not zero definitely and indicates classical chaos.

\medskip

Finally, it should be remarked that the finite-size effect of a string has played a significant role in generating classical chaos. The stringy ansatz (\ref{stringy}) reflects the finite size of a string while it has not been taken into account in the ansatz employed in \cite{Asano:2015qwa}. This observation is also consistent with the result of \cite{McLoughlin:2022jyt} in the Laudau-Lifshitz limit \cite{LL1,LL2}. It is because the spatial direction of the string world-sheet is taken to be infinite in this limit and the finite size effect of a string is dropped off. Hence classical chaos does not appear in this limit as discussed in \cite{McLoughlin:2022jyt}. On the other hand, the weak chaos in the next-nearest-neighbor interacting spin chain has been discussed before taking the thermodynamic limit and the finite-size effect has been included in this analysis.

\medskip

\newpage
\section{Conclusion and Discussion}

In this note, we have reconsidered string motion in a near pp-wave limit of AdS$_5\times $S$^5$ and shown the presence of classical chaos under a stringy ansatz by computing Poincar\'{e} sections and Lyapunov spectrum. As a result, the string theory in this limit is not classically integrable, while it is integrable both on the original AdS$_5\times $S$^5$ and the exact pp-wave background. This result is quite analogous to the Henon-Heiles model a a truncation of a periodic three particle Toda chain. 

\medskip 

Here we have studied classical chaos appearing in the middle process of the Penrose limit. What is the significance of this classical chaos? The near pp-wave geometry is just an approximation of the full background. No classical chaos appears in the string theory on the full AdS$_5\times$S$^5$ background as in the Toda chain. In this sense, this chaos may be seen as a perturbative artifact. 

\medskip

However, it might be possible to consider the gauge-theory counterpart by expanding the dilatation operator perturbatively as discussed in \cite{McLoughlin:2022jyt}. It would be interesting to check the AdS/CFT correspondence for the intermediate chaotic strings. The near pp-wave geometry is still a solution to type IIB supergravity \cite{Takayama}, on which the world-sheet beta function vanishes. It is also nice to try to reveal the relation between a near pp-wave string state and a truncated dilatation operator in the $\mathcal{N}=4$ SYM at quantum level. 

\medskip

We hope that our result would shed light on a new arena to study the AdS/CFT correspondence beyond the integrability.

\subsection*{Acknowledgments}

We are very grateful to Koji Hashimoto for useful comments and discussions. The work of S.K.\ was supported by JST SPRING, Grant Number JPMJSP2110. The work of K.Y.\ was supported by the Supporting Program for Interaction-based Initiative Team Studies (SPIRITS) from Kyoto University,  
JSPS Grant-in-Aid for Scientific Research (B) No.\,18H01214 and 22H01217, and MEXT KAKENHI Grant-in-Aid for Transformative Research Areas A “Extreme Universe” No.\ 22H05259 and ``Machine Learning Physics'' No.\ 22H05115.

\appendix

\section*{Appendix}

\section{The third conserved charge $I_3$}
\label{appendix} 

Let us present here the third conserved charge in a periodic three-particle Toda chain. 

\medskip

In order to figure out the speciality of the exponential potential in the Toda chain, it is instructive to start from more general setup as follows: 
\begin{align}
    H=\frac{1}{2}\qty(P_1^2+P_2^2+P_3^2)+V_1(Q_3-Q_2)+V_2(Q_1-Q_3)+V_3(Q_2-Q_1)\,. 
\end{align}
Here the potential $V_i~(i=1,2,3)$ are arbitrary real scalar potentials depending only on the difference of dynamical variables $Q_i$'s by supposing the translational invariance.    
A special case where 
\begin{eqnarray}
V_1=e^{-(Q_3-Q_2)}-1\,, \qquad V_2=e^{-(Q_1-Q_3)}-1\,, \qquad 
V_3=e^{-(Q_2-Q_1)}-1 \label{toda}
\end{eqnarray}
corresponds to the Toda chain (\ref{expHam})\,. 

\medskip 

Then, Hamilton's equations are
\begin{align}
  \dv{Q_1}{t}&=P_1\,,\qquad\dv{Q_2}{t}=P_2\,,\qquad\dv{Q_3}{t}=P_3\,,\label{geneom1}\\
  \dv{P_1}{t}&=-\qty(\pdv{V_2}{Q_1}+\pdv{V_3}{Q_1})\,,\quad\dv{P_2}{t}=-\qty(\pdv{V_1}{Q_2}+\pdv{V_3}{Q_2})\,,\quad\dv{P_3}{t}=-\qty(\pdv{V_1}{Q_3}+\pdv{V_2}{Q_3})\label{geneom2}\,.    
\end{align}
The total momentum and energy are conserved charge. 

\medskip 

It is the turn to consider the third conserved charge. Let us introduce the following quantity $I_3$~: 
\begin{align}
    I_3 \equiv P_1P_2P_3-P_1V_1-P_2V_2-P_3V_3\,.
\end{align}
Then it is useful to evaluate the time derivative of $P_1P_2P_3$ as 
\begin{align}
    \dv{t}\qty(P_1P_2P_3)&=+P_2P_3\dv{P_1}{t}+P_1P_3\dv{P_2}{t}+P_1P_2\dv{P_3}{t}\nonumber\\
    &=-P_2P_3\qty(\pdv{V_2}{Q_1}+\pdv{V_3}{Q_1})-P_1P_3\qty(\pdv{V_1}{Q_2}+\pdv{V_3}{Q_2})-P_1P_2\qty(\pdv{V_1}{Q_3}+\pdv{V_2}{Q_3})\nonumber\\
    &=P_2P_3\qty(\pdv{V_2}{Q_3}+\pdv{V_3}{Q_2})+P_1P_3\qty(\pdv{V_1}{Q_3}+\pdv{V_3}{Q_1})+P_1P_2\qty(\pdv{V_1}{Q_2}+\pdv{V_2}{Q_1})\nonumber\\
    &=P_1\dv{V_1}{t}+P_2\dv{V_2}{t}+P_3\dv{V_3}{t}\nonumber\\
    &=\dv{t}(P_1V_1+P_2V_2+P_3V_3)-\qty(\dv{P_1}{t}V_1+\dv{P_2}{t}V_2+\dv{P_3}{t}V_3)\,.
\end{align}
To make $I_3$ a conserved charge, the remaining part must vanish like 
\begin{align}
    \dv{P_1}{t}V_1+\dv{P_2}{t}V_2+\dv{P_3}{t}V_3=0\,.
\end{align}
By using Hamilton's equations in \eqref{geneom2}, it can be rewritten as 
\begin{align}
    \qty(\pdv{V_2}{Q_1}+\pdv{V_3}{Q_1})V_1+\qty(\pdv{V_1}{Q_2}+\pdv{V_3}{Q_2})V_2+\qty(\pdv{V_1}{Q_3}+\pdv{V_2}{Q_3})V_3=0\,.\label{veq}
\end{align}
Since the exponential potential (\ref{toda}) in the Toda chain satisfies \eqref{veq}\,, $I_3$ becomes the third conserved charge.
However, when the exponential potential is truncated to a finite order, the relation \eqref{veq} is not satisfied any more.


\end{document}